\documentclass[11pt]{revtex4-2}
\UseRawInputEncoding
\usepackage{bm}
\usepackage{amsmath}
\usepackage{amssymb}
\usepackage{graphicx}
\begin{document}
\title{New insights into four-boson renormalization group limit cycles}
\author{Bastian Kaspschak}
\affiliation{Helmholtz-Institut f\"ur Strahlen- und Kernphysik and Bethe Center for Theoretical Physics,
Universit\"at Bonn, D-53115 Bonn, Germany}
\author{Ulf-G. Mei{\ss}ner}
\affiliation{Helmholtz-Institut f\"ur Strahlen- und Kernphysik and Bethe Center for Theoretical Physics,
Universit\"at Bonn, D-53115 Bonn, Germany}
\affiliation{Institute for Advanced Simulation, Institut f\"ur Kernphysik, and J\"ulich Center
  for Hadron Physics, Forschungszentrum J\"ulich, D-52425 JĆ¼lich, Germany}
\affiliation{Tbilisi State University, 0186 Tbilisi, Georgia}

\date{\today}

\begin{abstract}
Using machine learning techniques, we verify that the emergence of renormalization group limit cycles 
beyond the unitary limit is transferred from the three-boson subsystems to the whole four-boson system.
Focussing on four identical bosons, we first
generate populations of synthetic singular potentials within the latent space of a boosted ensemble of 
variational autoencoders. After introducing the limit cycle loss for measuring the deviation of a given 
renormalization group flow from limit cycle behavior, we minimize it by applying an elitist genetic 
algorithm to the generated populations. 
The fittest potentials are observed to accumulate around the inverse-square potential, which 
we prove to generate limit cycles for four bosons and which is already known to produce limit cycles in 
the three-boson system. This also indicates that a four-body term does not enter low-energy observables
at leading order, since we do not observe any additional scale to emerge.
\end{abstract}

\maketitle

\section{Introduction}
Investigating $A$-body systems with unnaturally large scattering lengths has led to numerous
insights in the context of universality, see Refs.~\cite{BRAATEN2006259, PhysRevLett.118.202501,
  RevModPhys.89.035006,PhysRevLett.102.140401,doi:10.1146/annurev.nucl.52.050102.090637}. Low-energy
systems with resonant interactions share certain features independent of their short-range details.
Prominent examples of such universal features are the existence of a shallow dimer in the case $A=2$ or
the geometric accumulation of shallow trimers for $A=3$, also referred to as the Efimov effect,
see Refs.~\cite{EFIMOV1970563,osti_4068792}. The latter can be understood in terms of a
renormalization group (RG) limit cycle from which an additional scale arises.

Clearly, there is a deeper connection to renormalization and the construction of effective
field theories: A separation of scales can be leveraged to obtain an accurate low-energy description
of the system by integrating out all high-energy degrees of freedom beyond some short-range cutoff.
In order to renormalize the two-body system, a single two-body contact term suffices. However, due
to the Thomas collapse, see Ref.~\cite{PhysRev.47.903}, not only a two-body term, but also an
additional three-body term is required for the renormalization of the three-body system at leading
order (LO) in $l/|a|$. Here, $l$ and $a$ denote a typical low-energy length scale in the system and
the unnaturally large scattering length, respectively. There has been a lively debate on whether this
trend continues for larger numbers $A\geq 4$ of particles, that is if $A$-body terms appear at LO in
general. On the one hand, working with delta potentials, Ref.~\cite{PhysRevLett.74.487} argues
that adding an $A$-th particle in three spatial dimensions also introduces an additional ultraviolet
(UV) divergence to the kernel of the Lippmann-Schwinger equation. This would imply that a new scale
emerges and that an $A$-body term becomes necessary at LO. On the other hand, according to
Ref.~\cite{PhysRevD.7.2517}, the trace of the $A$-body kernel vanishes for $A\geq 4$. As this
cannot be compensated by a singularity in the connected $A$-body amplitude, there can only be
finitely many $A$-body bound states. The authors state that this makes the Efimov effect an
exclusive feature to the three-body system, which is the essence of the eponymous Amado-Greenwood
theorem. Thus, an additional scale does not arise and an $A$-body term is never introduced to LO.
There is even further evidence that supports this hypothesis: Trimer and tetramer energies seem to
be linearily correlated at LO. The resulting line in energy plane is known as the Tjon line,
see Ref.~\cite{TJON1975217} (strictly speaking, this is a band~\cite{Platter:2004zs}). In nuclear physics, one typically
considers ground state energies of the triton and $\alpha$ particle. Now the Tjon line indicates
that, after the three-body system has been renormalized, the tetramer energy can be predicted with
sufficient precision from the previously acquired two- and three-body parameters, again making a
four-body term at LO obsolete.

For the four-body system, Ref.~\cite{PhysRevA.70.052101} resolves this balancing act by showing that
once the two- and three-body subsystems have been renormalized, the tetramer energies are intrinsically
independent of the cutoff. Therefore, there is no need to introduce an additional four-body force,
since it does not appear in four-body observables at LO. Moreover, there is not only evidence on
correlations between trimer energies $E_3$ and tetramer energies $E_4$, but also pentamer and hexamer
energies $E_5$ and $E_6$ in the shape of generalized Tjon lines, see
Refs.~\cite{doi:10.1063/1.438195,PhysRevA.22.28,PhysRevA.94.052502}. Since these observations have
been made, $A$-body observables with $A\geq 4$ have been conjectured to depend only on the two-
and three-body parameters at LO as in Refs.~\cite{PhysRevA.94.052502,PhysRevLett.122.143001}.

At this point, it is also important to address the question which role the inverse square potential
plays in the $A$-body system. For $A=2$ and $A=3$ identical bosons interacting via pairwise inverse
square interactions are known to lead to an RG limit cycle, see
Refs.~\cite{PhysRevA.70.052111, HAMMER2006306, arXiv:2111.07820}. While universal properties of
$A$-body systems usually depend on resonant interactions, this is not the case for the mentioned
limit cycle behavior. In fact, we have shown that the inverse square potential is the only potential
to lead to an RG limit cycle in the three-body sector independent of the scattering length, see
Ref.~\cite{arXiv:2111.07820}. If the same could be observed for $A=4$ identical bosons, this
would allow us to make the following interesting implication when allowing the scattering length
to take arbitrary values: If any three-body subsystem gives rise to an RG limit cycle, the
four-body system automatically produces an RG limit cycle as well. This would deeply correspond
to the spirit of Ref.~\cite{PhysRevA.70.052101}, as predictions on the limit cycle aspect of
four-body physics can be made completely based on our knowledge about the three-body sector.
However, if the inverse square potential should either turn out to be no unique solution or no
solution at all in our search for four-body RG limit cycles, this would indicate an additional scale
making a four-body term necessary for renormalization. In that case, some of the previously mentioned
findings on the four-body system need to be revised.

Similar to our work on three-body RG limit cycles in Ref.~\cite{arXiv:2111.07820}, our starting point
is the low-energy Faddeev equation for $A=4$ identical bosons in hyperspherical coordinates. Due to
the fourth particle, an additional hyperangle appears. Assuming minimal node indices with respect to
both hyperangles allows us to solve the low-energy Faddeev equation locally and to connect these
local solutions by successive applications of generalized transfer matrices, see
Refs.~\cite{62122,arXiv:2111.07820}. Since we operate in configuration space, a delta-shell
regularization with respect to some hyperradial cutoff seems to be adequate. The computation
of the corresponding coupling constant is straightforward and involves an infinitesimal integration
over the low-energy Faddeev equation. Just as we did in Ref.~\cite{arXiv:2111.07820}, we define a
{\em limit cycle loss} (LCLoss) as a measure for log-periodicity: If the coupling constant, indeed,
depends log-periodically on the hyperradial cutoff, the vanishing LCLoss indicates the presence
of a limit cycle. Vice-versa, the larger the LCLoss becomes, the stronger the coupling constant deviates
from log-periodicity.

The search for limit cycles consists of running an elitist genetic algorithm (GA) based on
Goldberg's Simple Genetic Algorithm, see Ref.~\cite{goldberg1989genetic}, and favoring low
LCLosses. It is not directly carried out in the high-dimensional feature space of discretized and
finitely ranged two-body potentials, but in the low-dimensional latent space of the same boosted
ensemble of variational autoencoders (VAEs), see Ref.~\cite{kingma2013auto}, that has already been used
in Ref.~\cite{arXiv:2111.07820}, instead. This ensemble has been trained to encode and decode
two-body potentials from feature space and latent curves from latent space with sufficiently low
loss into each other. The advantage of this approach is not only a vigorous dimensionality reduction,
but also that it allows to generate synthetic two-body potentials from each point in latent space.
Especially the crossover and mutation steps of the applied GA thrive on this possibility to generate
synthetic potentials.

This paper is organized as follows:  
In Sec.~\ref{sec:four-bosons} we start with specifying the class of four-boson systems RG limit cycles will be searched for in this work. Basically, the local solutions of the low-energy Faddeev equation in hyperspherical coordinates are presented again, while relegating to App.~\ref{app:a} and \ref{app:b} for further technical details. Subsequently, we introduce generalized transfer matrices to connect these local solutions with each other in a continuously and differentiably way. Later on, we apply a delta-shell regularization on the resulting total wavefunctions, from which the coupling constants follow. These coupling constants, again, are highly relevant to the unsupervised search for RG limit cycles in Sec.~\ref{sec:unsupervised}, as they are an essential ingredient of the LCLoss introduced in this section. Sec.~\ref{sec:unsupervised} briefly explains the core aspects of our unsupervised search for RG limit cycles based on the latent space of the boosted VAE ensemble and later the elitist GA from Ref.~\cite{arXiv:2111.07820}. The fittest individuals resulting from the GA are analyzed in Ref.~\ref{sec:results}. The found accumulation point of fittest individuals in latent space is identified as the inverse-square potential, which is later proved to produce RG limit cycles in the four-boson sector for different hyperangle configurations.

\section{Four identical bosons at low energy}\label{sec:four-bosons}
\subsection{Low-energy wavefunction}
Our search for limit cycles is restricted to low-energy systems of four identical bosons with mass $m$ at the positions $\bm{r}_1,\bm{r}_2,\bm{r}_3,\bm{r}_4$ in coordinate space. We assume the total potential
\begin{equation}
V(\bm{r}_1,\bm{r}_2,\bm{r}_3,\bm{r}_4) = \sum\limits_{a=1}^3\sum\limits_{b>a}^4 v\left(|\bm{r}_b-\bm{r}_a|\right)
\end{equation}
to be a sum of finitely-ranged and piecewise-constant two-body potentials $v$. Each term $v\left(|\bm{r}_b-\bm{r}_a|\right)$ corresponds to the pairwise interaction between the $a$-th and the $b$-th particle.

Although there are twelve degrees of freedom for four identical bosons to deal with in general, this number can be drastically reduced by separating out the center-of-mass motion and, based on the low-energy criterion, by a projection onto S-wave states. The resulting equation of motion is the low-energy Faddeev equation for the four boson case, that is thoroughly motivated in App.~\ref{app:a}. With the Jacobi vectors
\begin{equation}
\bm{j}_{21}=\frac{1}{\sqrt{2}}\left(\bm{r}_2-\bm{r}_1\right), \ \bm{j}_{321}=\sqrt{\frac{2}{3}}\left(\bm{r}_3-\frac{\bm{r}_2+\bm{r}_1}{2}\right), \ \bm{j}_{4321}=\frac{\sqrt{3}}{2}\left(\bm{r}_4-\frac{\bm{r}_3+\bm{r}_2+\bm{r}_1}{3}\right)
\end{equation}
its remaining three degrees of freedom are covered by the hyperradius
\begin{equation}
R = \sqrt{j_{21}^2+j_{321}^2+j_{4321}^2}
\end{equation}
and the two hyperangles
\begin{equation}
\alpha = \arcsin\left(\frac{j_{321}}{\sqrt{j_{21}^2+j_{321}^2}}\right), \ \beta = \arcsin\left(\frac{j_{4321}}{\sqrt{j_{21}^2+j_{321}^2+j_{4321}^2}}\right). \label{eq:def-hyperangles-x1}
\end{equation}
These three hyperspherical coordinates are bound to $R\in[0,+\infty]$ and ${\alpha,\beta\in[0,\pi/2]}$, respectively. The distance between the first two bosons is given by
\begin{equation}
r=\sqrt{2}R\cos\alpha\cos\beta \label{eq:radius-hyperradius}
\end{equation}
and approaches the maximum value $r=\sqrt{2}R$ at vanishing hyperangles $\alpha=0$ and $\beta=0$. In App.~\ref{app:b} we consider the low-energy Faddeev equation in a region, where the two-body potential takes the constant value
\begin{equation}
v(\sqrt{2}R\cos\alpha\cos\beta)=\frac{u}{2m}.
\end{equation}
Neglecting hyperangular dependencies to the wavefunction in vicinity of $r=\sqrt{2}R\cos\alpha\cos\beta$ reduces the low-energy Faddeev equation to a purely hyperradial equation of motion. The resulting zero-node solutions
\begin{equation}
\Psi_{k_u}(R) = \frac{1}{R^{7/8}}\left(A J_{7/2}(k_u R)+B Y_{7/2}(k_u R)\right) \label{eq:local-solution}
\end{equation}
with momenta $k_u$ defined by
\begin{equation}
k_u^2 = 2mE-6u
\end{equation}
represent the local behavior of the total wavefunction as a linear combination of Bessel functions of first and second kind with purely hyperradial arguments times the monomial $R^{-7/8}$.

\subsection{Transfer matrices}
The fact that the considered two-body potential is piecewise-constant leads to solutions that behave locally like $\Psi_{k_u}$ in Eq.~\eqref{eq:local-solution}. Those radii $r_i$ that serve as boundaries of the constant potential steps, that is $v(r_{i-1}\leq r < r_i)= u_i/(2m)$, are referred to as transition radii. Using Eq.~\eqref{eq:radius-hyperradius}, we can translate transition radii into transition hyperradii,
\begin{equation}
R_i(\alpha,\beta) = \frac{r_i}{\sqrt{2}\cos\alpha \cos\beta},
\end{equation}
which can be understood as surfaces in $\mathbb{R}^3$.
From our previous findings we deduce two features of the global behavior of the total wavefunction: Locally, it behaves like a superposition as in Eq.~\eqref{eq:local-solution}. However, the factor $k^{(i)}=k_{u_i}$ only appears in the arguments of the Bessel functions if we consider the wavefunction within the specific segment between two adjacent surfaces $R_{i-1}(\alpha,\beta)$ and $R_i(\alpha,\beta)$. In order to obtain a smooth wavefunction, solutions of adjacent segments need to be connected with each other in a continuous and differentiable manner. This connection is achieved by the application of transfer matrices, see Ref.~\cite{62122}. At this point, we closely follow the approach of our work in Ref.~\cite{arXiv:2111.07820}, where we have thoroughly derived transfer matrices for the three-boson case. Similar to these transfer matrices depending also on the hyperangle $\alpha$, we expect transfer matrices in the four-boson case to depend on both hyperangles $\alpha$ and $\beta$. This also introduces hyperangular dependencies to the coefficients $A$ and $B$ in Eq.~\eqref{eq:local-solution}, which finally leads to a non-trivial hyperangular behavior of the total wavefunction.

At each surface $R_i(\alpha,\beta)$ the continuity and differentiability criteria become manifest in the following two equations that connect Bessel coefficients of adjacent segments with each other:
\begin{equation}
\begin{aligned}
&A^{(i)}(\alpha,\beta)\ J_{7/2}\left[k^{(i)} R_i(\alpha,\beta)\right]+B^{(i)}(\alpha,\beta)\ Y_{7/2}\left[k^{(i)} R_i(\alpha,\beta)\right] \\ &= A^{(i+1)}(\alpha,\beta)\ J_{7/2}\left[k^{(i+1)} R_i(\alpha,\beta)\right]+B^{(i+1)}(\alpha,\beta)\ Y_{7/2}\left[k^{(i+1)} R_i(\alpha,\beta)\right]÷,
\end{aligned}
\label{eq:continuity}
\end{equation}
and differentiability 
\begin{equation}
\begin{aligned}
&A^{(i)}(\alpha,\beta)k^{(i)}\ J_{7/2}^\prime\left[k^{(i)} R_i(\alpha,\beta)\right]+B^{(i)}(\alpha,\beta)k^{(i)}\ Y_{7/2}^\prime\left[k^{(i)} R_i(\alpha,\beta)\right] \\ &= A^{(i+1)}(\alpha,\beta)k^{(i+1)}\ J_{7/2}^\prime\left[k^{(i+1)} R_i(\alpha,\beta)\right]+B^{(i+1)}(\alpha,\beta)k^{(i+1)}\ Y_{7/2}^\prime\left[k^{(i+1)} R_i(\alpha,\beta)\right].
\end{aligned}
\label{eq:differentiability}
\end{equation}
Eqs.~\eqref{eq:continuity} and \eqref{eq:differentiability} can be combined to a single vector equation
\begin{equation}
\begin{pmatrix} A^{(i+1)}(\alpha,\beta) \\ B^{(i+1)}(\alpha,\beta) \end{pmatrix} = \mathrm{T}^{(i)}(v,  \alpha, \beta) \begin{pmatrix} A^{(i)}(\alpha,\beta) \\ B^{(i)}(\alpha,\beta) \end{pmatrix}÷.
\label{eq:transfer}
\end{equation}
Given the coefficients $A^{(i)}(\alpha,\beta)$ and $B^{(i)}(\alpha,\beta)$ in the $i$-th segment, therefore, suffices to deduce the Bessel coefficients $A^{(i+1)}(\alpha,\beta)$ and $B^{(i+1)}(\alpha,\beta)$ in the subsequent segment. In Eq.~\eqref{eq:transfer}, the actual connection between segments is established by the transfer matrix $\mathrm{T}^{(i)}(v,  \alpha, \beta)$. With the product
\begin{equation}
\begin{aligned}
\left\{f, g\right\}^{(i)}(v,  \alpha, \beta) =\ &k^{(i+1)}f_{7/2}\left[k^{(i)} R_i(\alpha, \beta)\right]g_{5/2}
\left[k^{(i+1)} R_i(\alpha, \beta)\right]\\ &-k^{(i)}f_{5/2}\left[k^{(i)} R_i(\alpha, \beta)\right]g_{7/2}
\left[k^{(i+1)} R_i(\alpha, \beta)\right]
\end{aligned}
\end{equation}
for two families $\{f_x\}_{x\in\mathbb{R}}$ and $\{g_x\}_{x\in\mathbb{R}}$ of differentiable functions, the transfer matrix between the $i$-th and the $(i+1)$-th segment can be written as
\begin{equation}
\mathrm{T}^{(i)}(v,  \alpha, \beta) = \frac{\pi}{2}R_i(\alpha,\beta)\ \begin{pmatrix} +\left\{J, Y\right\}^{(i)}(v,  \alpha, \beta) & +\left\{Y, Y\right\}^{(i)}(v,  \alpha, \beta) \\ -\left\{J, J\right\}^{(i)}(v,  \alpha, \beta)
& -\left\{Y, J\right\}^{(i)}(v,  \alpha, \beta) \end{pmatrix}.
\label{eq:matrices}
\end{equation}
By a successive application of transfer matrices, it is possible to infer the Bessel coefficients $A^{(i)}(\alpha,\beta)$ and $B^{(i)}(\alpha,\beta)$ in the $i$-th segment from those in the very first segment, that is $A^{(1)}(\alpha,\beta)$ and $B^{(1)}(\alpha,\beta)$. Since the Bessel function $Y_{7/2}(x)$ is singular at $x=0$, the corresponding coefficient in the first segment needs to vanish in order to ensure square-integrability, that is $B^{(1)}(\alpha,\beta)=0$. The entire normalization of the total wavefunction, therefore, depends on the value of $A^{(1)}(\alpha,\beta)$. Without loss of generality, we choose $A^{(1)}(\alpha,\beta)=1$, which also fixes all remaining Bessel coefficients:
\begin{equation}
A^{(i)}(\alpha,\beta) = \begin{pmatrix} 1, & 0 \end{pmatrix}\left( \prod\limits_{j=1}^{i-1}
 T^{(j)}(v, \alpha,\beta)\right) \begin{pmatrix} 1 \\ 0 \end{pmatrix}÷,
\end{equation}
\begin{equation}
B^{(i)}(\alpha,\beta) = \begin{pmatrix} 0, & 1 \end{pmatrix}\left( \prod\limits_{j=1}^{i-1}
 T^{(j)}(v, \alpha,\beta)\right) \begin{pmatrix} 1 \\ 0 \end{pmatrix}÷.
\end{equation}

\subsection{Delta-shell regularization}
Similar to our work in Ref.~\cite{arXiv:2111.07820}, the piecewise-constant two-body potentials approximate singular potentials, which are known to produce unphysical infinities in the short-range sector. At this point, Wilsonian renormalization comes into play: The key idea is to eliminate these infinities by integrating out the responsible short-range degrees of freedom below some short-range cutoff. Since low-energy observables need to be independent of the cutoff, counterterms are introduced, which correspond to additional couplings.

We again apply delta-shell regularization to two-body potentials. This means that $v(\sqrt{2}\cos\alpha\cos\beta)$ is substituted by
\begin{equation}
v(R_*|\ R,\alpha,\beta) = \begin{cases}-\displaystyle\frac{1}{4mR_*\cos^2\alpha\cos^2\beta} H(R_*,\alpha,\beta)\delta(R-R_*) & R<R_* \\
+v(\sqrt{2}R\cos\alpha\cos\beta) & R\geq R_* 
\end{cases},
\label{eq:deltashell}
\end{equation}
where $R_*=r_*/(\sqrt{2}\cos\alpha\cos\beta)$ is the cutoff hyperradius that follows from a given cutoff radius $R_*$ and hyperangle pair $\alpha$ and $\beta$. Note that delta-shell regularization does not affect the two-body potential beyond the cutoff hyperradius. The actual substitution takes place for hyperradii $R\in[0,R_*)$ instead. Here $v(R_*|\ R,\alpha,\beta)$ only takes non-zero values infinitesimally close to the delta-shell with radius $R_*$. Note that by successively increasing the cutoff hyperradius, increasingly more short-range degrees of freedom of the original two-body potential are eliminated.

The coupling constant $H(R_*,\alpha,\beta)$ is essential to our search for limit cycles in the four-boson case. Its purpose is to ensure cutoff-independence of low-energy observables by satisfying a corresponding matching condition. Quantities that are usually considered in this context are either tetramer ground-state energies or, as we will do in the following, the zero-energy wavefunction. That means we derive the wavefunction for the regularized four-boson system and adjust the coupling constant $H(R_*,\alpha,\beta)$ such that it equals the wavefunction of the unregularized system beyond the hyperradial cutoff.

Since we have already presented a detailed derivation of the coupling constant in Ref.~\cite{arXiv:2111.07820}, we only highlight the most important aspects in the four-body case: Due to the delta function in Eq.~\eqref{eq:deltashell}, $H(R_*,\alpha,\beta)$ can be obtained by an infinitesimal hyperradial integration over the low-energy Faddeev equation (see App.~\ref{app:a}),
\begin{equation}
\begin{aligned}
H(R_*,\alpha, \beta)&= -4mR_*\cos^2\alpha\cos^2\beta\int_{R_*-\delta R}^{R_*+\delta R} \mathrm{d}R \, v(R_*|\ R,\alpha,\beta)\\
&=\frac{2}{3}mR_*\cos^2\alpha\cos^2\beta\int_{R_*-\delta R}^{R_*+\delta R} \mathrm{d}R \, \frac{1}{\Psi(R,\alpha,\beta)}(T_R+T_\alpha+T_\beta-E)
\Psi(R,\alpha,\beta).
\label{eq:infinitesimalintegral}
\end{aligned}
\end{equation}
Due to the infinitesimal integration around the delta-shell, the contribution of most integrands in Eq.~\eqref{eq:infinitesimalintegral} will vanish. It is only the second-order hyperradial derivatives originating from the hyperradial kinetic energy operator 
\begin{equation}
T_R=-\frac{1}{2m}\left(\frac{\partial}{\partial R^2}+\frac{8}{R}\frac{\partial}{\partial R}\right)
\end{equation}
that are allowed to be discontinuous and that produce non-zero contributions to the coupling constant. Resolving the integral yields
\begin{equation}
\begin{aligned}
H(R_*,\alpha, \beta) &= -\frac{1}{3}R_*\cos^2\alpha\cos^2\beta\int_{R_*-\delta R}^{R_*+\delta R} \mathrm{d}R \, \frac{\Psi^{\prime\prime}(R,\alpha,\beta)}{\Psi(R,\alpha,\beta)}\\
&=\frac{1}{3}R_*\cos^2\alpha\cos^2\beta\left[
\lim_{R\to R_*^-}\frac{\Psi^{(1)\prime}(R,\alpha,\beta)}{\Psi^{(1)}(R,\alpha,\beta)}-\lim_{R\to R_*^+}
\frac{\Psi^{(2)\prime}(R,\alpha,\beta)}{\Psi^{(2)}(R,\alpha,\beta)}\right],
\label{eq:difflogderivs}
\end{aligned}
\end{equation}
where $\Psi^{(1)}(R,\alpha,\beta)$ denotes the solution of the effectively free four-boson system with $R<R_*$ inside of the delta-shell and $\Psi^{(2)}(R,\alpha,\beta)$ denotes the wavefunction outside of the delta-shell. Let us assume that the cutoff hyperradius is located in the $i$-th segment, that is 
\begin{equation}
R_{i-1}(\alpha,\beta)\leq R_*(\alpha,\beta) < R_{i}(\alpha,\beta).
\end{equation}
Inserting Eq.~\eqref{eq:local-solution} with $k_u=\sqrt{-6u}$, where $u$ vanishes inside of the delta-shell, into Eq.~\eqref{eq:difflogderivs} and approximating Bessel functions with their asymptotic behavior at small arguments as in Ref.~\cite{arXiv:2111.07820} then yields the coupling constant
\begin{equation}
H(R_*,\alpha,\beta)=\frac{1}{3}\cos^2\alpha\cos^2\beta
\left[7-k^{(i)} R_*\frac{A^{(i)}(\alpha,\beta) J_{5/2}[k^{(i)} R_*] + B^{(i)}(\alpha,\beta) Y_{5/2}[k^{(i)} R_*]}
{A^{(i)}(\alpha,\beta) J_{7/2}[k^{(i)} R_*] + B^{(i)}(\alpha,\beta) Y_{7/2}[k^{(i)} R_*]}\right].
\label{eq:hcoupling}
\end{equation}

\section{Unsupervised search for RG limit cycles}\label{sec:unsupervised}
Now that we know how to compute the coupling constant $H(R_*,\alpha,\beta)$ given a pair $\alpha,\beta$ of hyperangles, a cutoff hyperradius $R_*$ and Bessel coefficients $A^{(i)}(\alpha,\beta), B^{(i)}(\alpha,\beta)$ from a transfer matrix approach in each segment, we can turn to the search for RG limit cycles. More precisely, we search for two-body potentials $v$ that cause $H(R_*,\alpha,\beta)$ to behave log-periodically in terms of $R_*$. As this corresponds to the desired limit cycle behavior, we will refer to such potentials as being LC for the sake of brevity.
\subsection{Synthetic two-body potentials}
Strictly speaking, a search for LC potentials as described above would have to be performed in an infinitely dimensional function space. Given an objective function that measures the deviation of $H(R_*,\alpha,\beta)$ from log-periodicity and, therefore, needs to be minimized, this search can be understood as a non-linear optimization problem in infinitely many dimensions. However, insisting on this naive approach is neither computationally feasible nor especially insightful. Instead, we can leverage the fact that the scope of Sec.~\ref{sec:four-bosons} has been restricted to piecewise-constant and finitely-ranged potentials: These two properties allow us to treat our two-body potentials as vectors $\bm{u}$, where $u_i/(2m)$ is the value $v(\sqrt{2}R\cos\alpha\cos\beta)$ takes for $R_{i-1}\leq R < R_i$. In the following we will use $\bm{u}$ and $u(r)=2mv(r)$ synonymously. We see that each component $u_i$, which we will also call the $i$-th feature of $\bm{u}$ in order to stick to machine learning terminology, is directly related to a specific potential step. By construction, the number $F=10^3$ of fixed transition radii
\begin{equation}
r_i = \exp\left( 12 \frac{i-F}{F-1}\right)
\end{equation}
is also the number of features within $\bm{u}$. However, the effective dimension of the search space is even lower than $F$. This is especially due to the fact that all considered potentials are attractive and approximate a singulatrity in the origin. A typical estimate of that effective dimension is the latent space dimension $L$ of autoencoders, see Refs.~\cite{bourlard1988auto,10.5555/2987189.2987190}, that faithfully reproduce given inputs: Their bottleneck architecture enforces a severe information loss during encoding, such that the decoder needs to reconstruct the usually high-dimensional input from a much lower-dimensional latent vector containing only the $L$ most distinctive features. When using variational autoencoders (VAEs), see Ref.~\cite{kingma2013auto}, it is even possible to generate new potentials from latent space. Since these potentials do not appear in the original training and test data sets, they are also referred to as synthetic potentials.

At this point it is obvious to use the pretrained models from Ref.~\cite{arXiv:2111.07820}, since the former search space coincides with the current search space for four-boson limit cycles. Back then, we used a boosted ensemble $\mathcal{A}=\mathcal{D}\circ\mathcal{E}$ of $N_C=8$ variational autoencoders $\mathcal{A}_i=\mathcal{D}_i\circ \mathcal{E}_i$ with $i=1,\ldots, 8$, where $\mathcal{E}_i:\mathbb{R}^F\to\mathbb{R}^L$ and $\mathcal{D}_i:\mathbb{R}^L\to\mathbb{R}^F$ are the $i$-th encoder and decoder, respectively, that either map to or from an $L=3$ dimensional latent space. Since the individual latent spaces of the $\mathcal{A}_i$ are independent from each other, the effective latent space dimension of the whole ensemble is $N_C\times L=24$. 

\begin{figure}[t]
\includegraphics[width=0.66\textwidth]{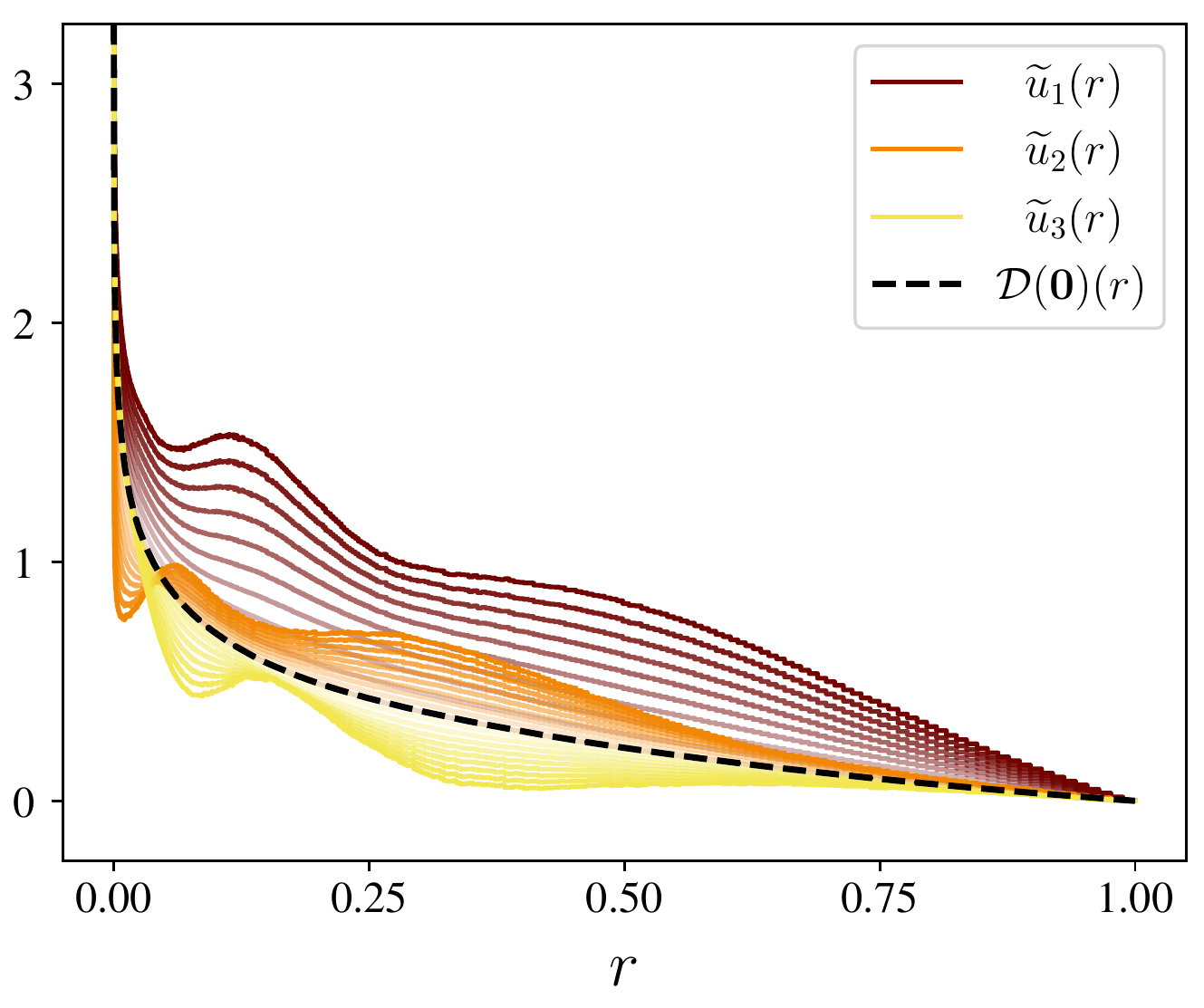}
\caption{The dashed line represents the decoded origin of the latent space that has been established by the boosted VAE ensemble $\mathcal{A}$ in Ref.~\cite{arXiv:2111.07820}. Due to its central position in latent space, its shape can be understood as the average potential shape of the data set used for training. By iteratively moving away from the latent space origin with a constant step size along three random directions and decoding each attained point, three different families of synthetic two-body potentials emerge. Here, opacity visualizes the distance between the respective point and the latent space origin.}
\label{fig:SyntheticPotentials}
\end{figure}

It must be mentioned that prior to the training pipeline, all potentials have been transformed to have all features of the same order of magnitude among the whole training set. Given the raw potential $\bm{u}$, this non-linear transformation yields potentials with the features
\begin{equation}
\widetilde{u}_i=\frac{1}{8}\log(-u_i). \label{eq:transformed-potential}
\end{equation}
Consequently, $\mathcal{A}$ actually encodes the transformed potentials $\widetilde{\bm{u}}$, which is especially important when dealing with synthetic potentials. Fig.~\ref{fig:SyntheticPotentials} shows several synthetic potentials as provided by $\mathcal{D}$. Due to being transformed, an inverse transformation has to be implemented at the end of the decoding pipeline. Finally, it is the potentials $\bm{u}$ with features $u_i=-\exp(8\widetilde{u}_i)$ resulting from the inverse transformation that can be used to derive RG flows for synthetic potentials. In fact, both data sets used to train and validate the ensemble~$\mathcal{A}$ are entirely composed of such transformed potentials. Further information on training and architecture of the $\mathcal{A}_i$ as well as on the effective encoder $\mathcal{E}$ and $\mathcal{D}$ can be found in Ref.~\cite{arXiv:2111.07820}.

\subsection{Limit cycle loss and genetic algorithm}
Using the pretrained ensemble $\mathcal{A}$ from Ref.~\cite{arXiv:2111.07820}, we have reduced the search for LC potentials in high-dimensional feature space to an optimization problem in low-dimensional latent space. Each point in latent space corresponds to a two-body potential for which the low-energy Faddeev equation can be solved and the RG flow of the coupling constant $H(R_*,\alpha,\beta)$ can be derived. The decisive criterion by which we determine LCness is the log-periodicity of $H(R_*,\alpha,\beta)$ in the interval
\begin{equation}
P = \left[ \log(-12) - \log(\sqrt{2}\cos\alpha\cos\beta) , - \log(\sqrt{2}\cos\alpha\cos\beta) \right].
\end{equation}
This also needs to be reflected in the objective function we want to minimize. Again we work with the limit cycle loss (LCLoss) that has already been used to identify RG limit cycles for three identical bosons in Ref.~\cite{arXiv:2111.07820}. Once we know $H(R_*,\alpha,\beta)$ for some pair of fixed hyperangles $\alpha$ and $\beta$ we first need to find those cutoff hyperradii $p_i$ at which it becomes singular. Log-periodicity requires the logarithms of these cutoff hyperradii to be equidistant. Therefore, we search for two-body potentials that cause the distribution $\Pi$ of all distances $p_{i+1}-p_i$ between adjacent poles two have a minimum standard deviation. The LCLoss itself is again defined as the quotient of the standard deviation and mean of $\Pi$,
\begin{equation}
\mathcal{L}_\text{LC} = \frac{\sigma(\Pi)}{\mu(\Pi)}.
\end{equation}
Note that in the limit of perfect log-periodicity, $\sigma(\Pi)$ and, consequently, $\mathcal{L}_\text{LC}$ vanish. However, deviations from log-periodicity increase $\mathcal{L}_\text{LC}$ and are penalized during optimization.

\begin{figure}[t]
\includegraphics[width=0.66\textwidth]{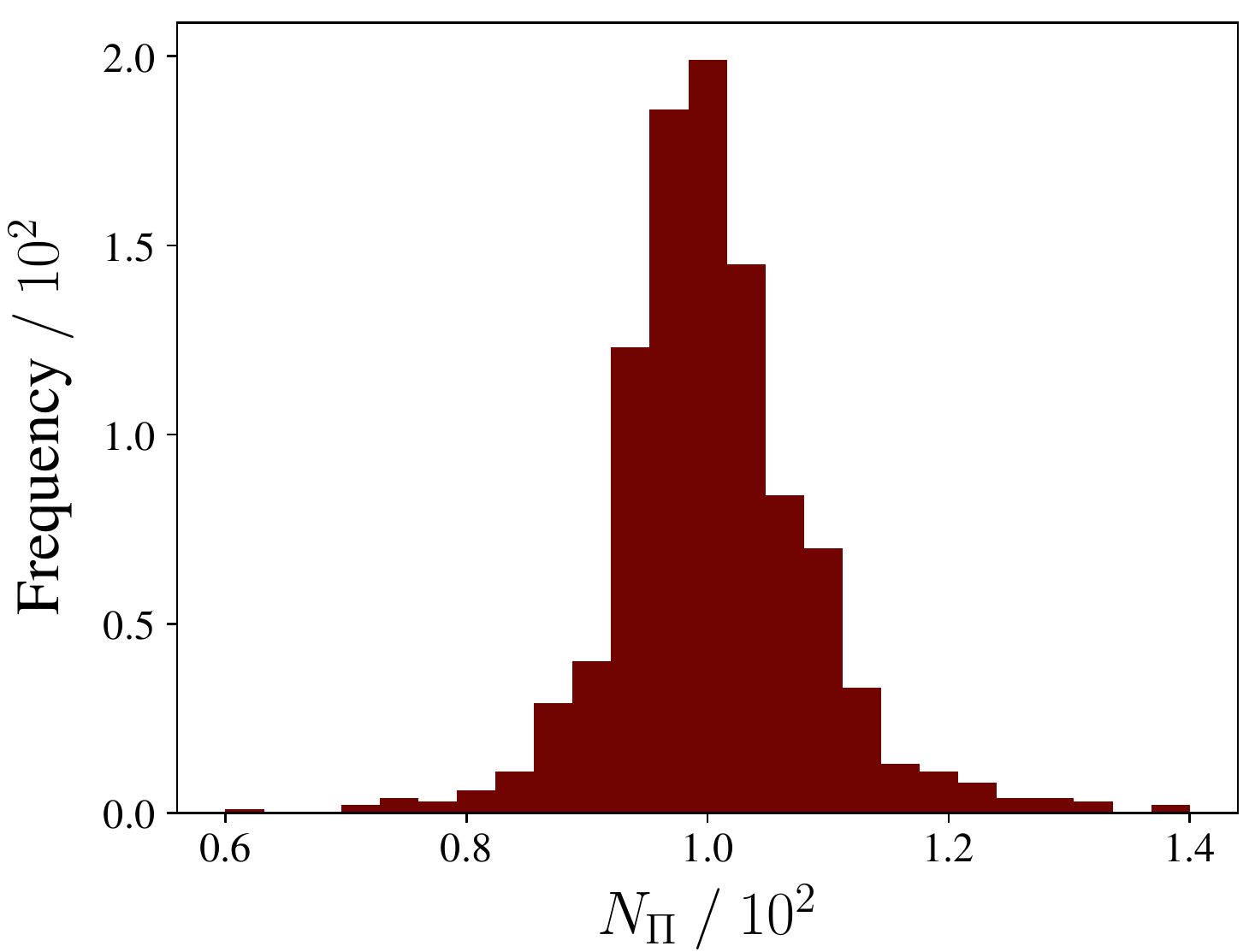}
\caption{Distribution of the number $N_\Pi$ of poles of the coupling constant $H(R_*,\alpha,\beta)$ for $10^3$ synthetic potentials randomly drawn from the standard distribution in latent space. From these $10^3$ synthetic potentials, $19$ are not represented in the shown histogram, since their values for $N_\pi$ exceed the target value of $N_\Pi=100$ by almost one order of magnitude. These outliers correspond to points that are comparatively remote from the latent space origin and cause the standard deviation of the whole $N_\Pi$ distribution to take the notably large value of $\sigma=61.0$. However, dropping these outliers and only considering entries in the displayed interval, we obtain a satisfactorily small standard deviation of $\sigma=8.2$ and a mean of $\mu=100.3$, which is much closer to the target value.}
\label{fig:HistogramNumPoles}
\end{figure}

An important aspect of the RG flow of $H(R_*,\alpha,\beta)$ that must not be ignored during the computation of LCLosses is that the number $N_\Pi=|\Pi|+1$ 
of poles in the interval $P$ with $|\Pi|$ being the number of elements in $\Pi$ sensitively depending on the depth of the considered potential. The deeper a potential, the more poles are encountered by the RG flow of $H(R_*,\alpha,\beta)$. While the LCLoss cannot be reliably estimated if there are only $N_\Pi\sim \mathcal{O}(1)$ poles, there is the risk that the pole frequency exceeds the hyperradial resolution of $10^5$ for extremely deep potentials. Let us consider scaling by a constant factor: With the scaled potential
\begin{equation}
u_i = -\exp[8(\widetilde{u}_i+s)]
\end{equation}
this scaling operation can also be understood as an addition of a constant term $s$ to all features of the transformed potential. It plays a special role because it never makes a non-LC potential LC or vice-versa, see Ref.~\cite{arXiv:2111.07820}. Therefore, it is sufficient to scale all synthetic potentials to a reasonable depth at which none of the two mentioned issues occurs. Without loss of generality, we again choose $s$ such that $N_\Pi$ approximately takes the value $100$ at the hyperangles $\alpha=0$ and $\beta=0$. In order to scale synthetic potentials appropriately, we supervisedly train a boosted ensemble $\mathcal{S}$ of CNNs with the same architecture as in Ref.~\cite{arXiv:2111.07820} to predict scales $s$ for given inputs in latent space. The inputs in the training and test data sets are again drawn from a standard distribution in latent space. The only difference to our approach in Ref.~\cite{arXiv:2111.07820} is that the scales themselves are no longer obtained by a grid search, but by a bisection in the range $s\in[-15,+5]$. The bisection is terminated if the maximum of $20$ iterations is reached or if a value $s$ is found such that $H(R_*,\alpha,\beta)$ has exactly $N_\Pi=100$ poles for $R_*\in P$. When inspecting $10^3$ synthetic potentials drawn from a standard distribution in latent space, we find the distribution of $N_\Pi$ to have the main $\mu=105.4$ and standard deviation $\sigma=61.0$. Compared to the optimal value of $N_\Pi=100$, this is a rather large error. However, this is mainly due to very few outliers with pole numbers of the order $N_\Pi\sim \mathcal{O}(10^3)$, which also correspond to outliers of the latent space distribution. When restricting the pole number to $N_\Pi\in[60, 140]$, only $19$ entries are dropped, while $981$ entries remain. Their distribution is shown in Fig.~\ref{fig:HistogramNumPoles} and has its mean $\mu=100.3$ much closer to the optimal value as well as a much lower standard deviation of $\sigma = 8.2$.

As a map from latent space to $\mathbb{R}_+$, the LCLoss has a notably complex loss surface topology. Due to scaling potentials such that the coupling constant $H(R_*,\alpha,\beta)$ produces approximately $N_\Pi=100$ poles, we obtain a reliable estimate on the LCLoss that mainly depends on the shape and not on the depth of the synthetic potentials. Nonetheless, the loss surface still produces a minor bump whenever a pole enters or leaves the interval $P$. In addition, the global behavior is highly non-convex. Therefore, gradient-descent techniques are ruled out and we require another optimization approach.

We decide to continue exactly as in Ref.~\cite{arXiv:2111.07820}, where an elitist Genetic Algorithm similar to Goldberg's Simple Genetic Algorithm (SGA) is applied to $50$ populations of $100$ synthetic potentials each over $100$ generations in parallel. Relegating to Ref.~\cite{arXiv:2111.07820} for technical details, we want to recapitulate the main principle of GAs: Given a measure of fitness (the LCLoss) for all individuals (the synthetic potentials), the key idea of GAs is that fitter individuals are favored during crossover selection and, therefore, more likely to produce offspring than the remaining, weaker individuals. The offspring again shares genes (latent space coordinates) of both parents and experiences mutation, by which it is possible to further explore the fitness landscape, or equivalently the loss surface, and even extend the domain of the initial population. Each generation consists of crossover selection, crossover, mutation and death selection. While weaker individuals will be dropped over time, increasingly fitter individuals are produced and ideally accumulate around well performing local minima or even the global minimum of the loss surface. In this particular case, we can infer accumulation points from analyzing the fittest individuals from each population after $100$ generations and identify them with the desired LC potentials. 
\begin{figure}[t]
\includegraphics[width=0.66\textwidth]{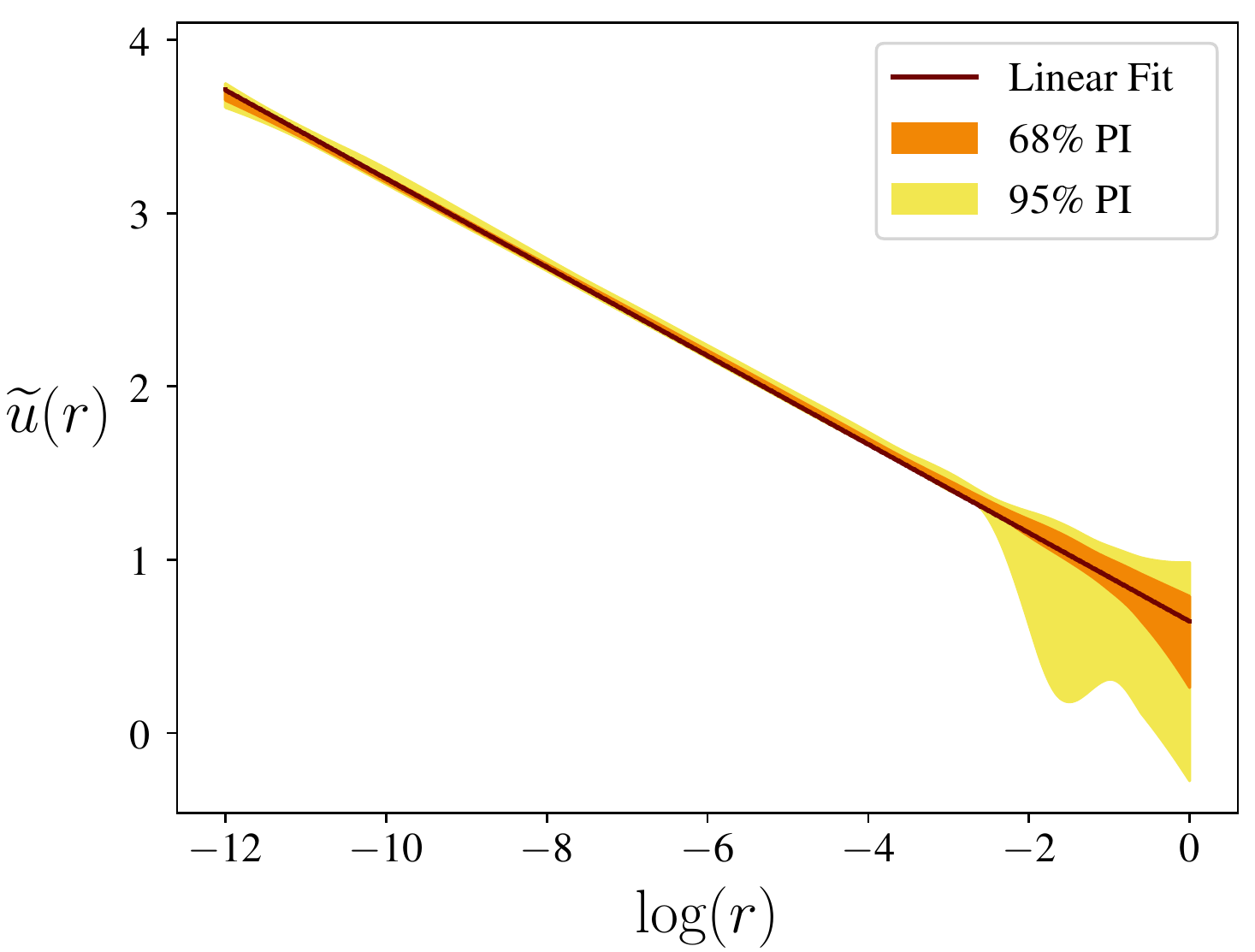}
\caption{The fifty fittest individuals resulting from the elitist GA are fitted with a linear function
$\widetilde{u}_\text{fit}(r)=a\log(r)+s$.
All fittest individuals behave strikingly similar in the short range regime with $\log(r)\lesssim -3$. However, as it is reflected in the 68\% and 95\% prediction intervals (PI) of the distribution of fittest individuals, there is a broadening in the long range regime $\log(r)\gtrsim -3$ due to outliers.}
\label{fig:LinearFitAndPredictionIntervals}
\end{figure}

\section{Results}\label{sec:results}
Fig.~\ref{fig:LinearFitAndPredictionIntervals} displays the $68\%$ and $95\%$ prediction intervals of the distribution that contains all fifty fittest individuals as well as a linear fit $\widetilde{u}_\text{fit}(r)=a\log(r)+s$ through all data points. In the short-range regime with $\log(r)\lesssim -3$ the obtained synthetic potentials are remarkably similar, which follows from the narrow prediction bands. This sufficiently verifies that these individuals only accumulate around one point in latent space. It is also evident that this accumulation point actually corresponds to an LC potential, since the absolute fittest individual, which is close to the displayed fitted line in Fig.~\ref{fig:LinearFitAndPredictionIntervals}, has a satifactorily low LCLoss of $\mathcal{L}_\text{LC}=1.436\times 10^{-2}$. In Ref.~\cite{arXiv:2111.07820}, we found one unique accumulation point in latent space, which could be identified with the inverse square potential 
\begin{equation}
u(r) = -\frac{\exp(8s)}{r^2}, \label{eq:inverse-square-x1}
\end{equation}
or, when applying the transformation in Eq.~\eqref{eq:transformed-potential},
\begin{equation}
\widetilde{u}(r)=-\frac{1}{4} \log(r) + s. \label{eq:transformed-inverse-square-potential}
\end{equation}
Out of the two parameters $a_\text{fit} = (-2.555\pm 0.135)\times 10^{-1}$ and $s_\text{fit}=(6.453\pm 1.027)\times 10^{-1}$ of the fitted line in Fig.~\ref{fig:LinearFitAndPredictionIntervals}, we are particularly interested in $a_\text{fit}$. Since the slope $a=-1/4$ from Eq.~\eqref{eq:transformed-inverse-square-potential} deviates by less than $1\sigma$ from the fitted slope, this is, in fact, the same accumulation point that was already identified in Ref.~\cite{arXiv:2111.07820}. We can also convince ourselves of that conclusion by directly applying the inverse transformation to the fitted line, which yields the potential $u_\text{fit}(r)\propto 1/r^{p_\text{fit}}$ with the exponent
\begin{equation}
p_\text{fit}=-8a_\text{fit}=2.044\pm 0.108. \label{eq:exponent-x1}
\end{equation}
With $\mathcal{L}_\text{LC}=1.288\times 10^{-1}$, the LCLoss of the fitted line turns out to be much higher that the absolute fittest individual. However, this is most likely due to the influence of the outliers on the fitted slope. Substituting the fitted slope of the line with the value $a=-1/4$ yields the LCLoss $\mathcal{L}_\text{LC}=3.872\times 10^{-5}$ of the inverse-square potential, which is significantly lower and, thus, serves as additional evidence for this accumulation point to be an LC potential.

\begin{figure}[t]
\includegraphics[width=0.66\textwidth]{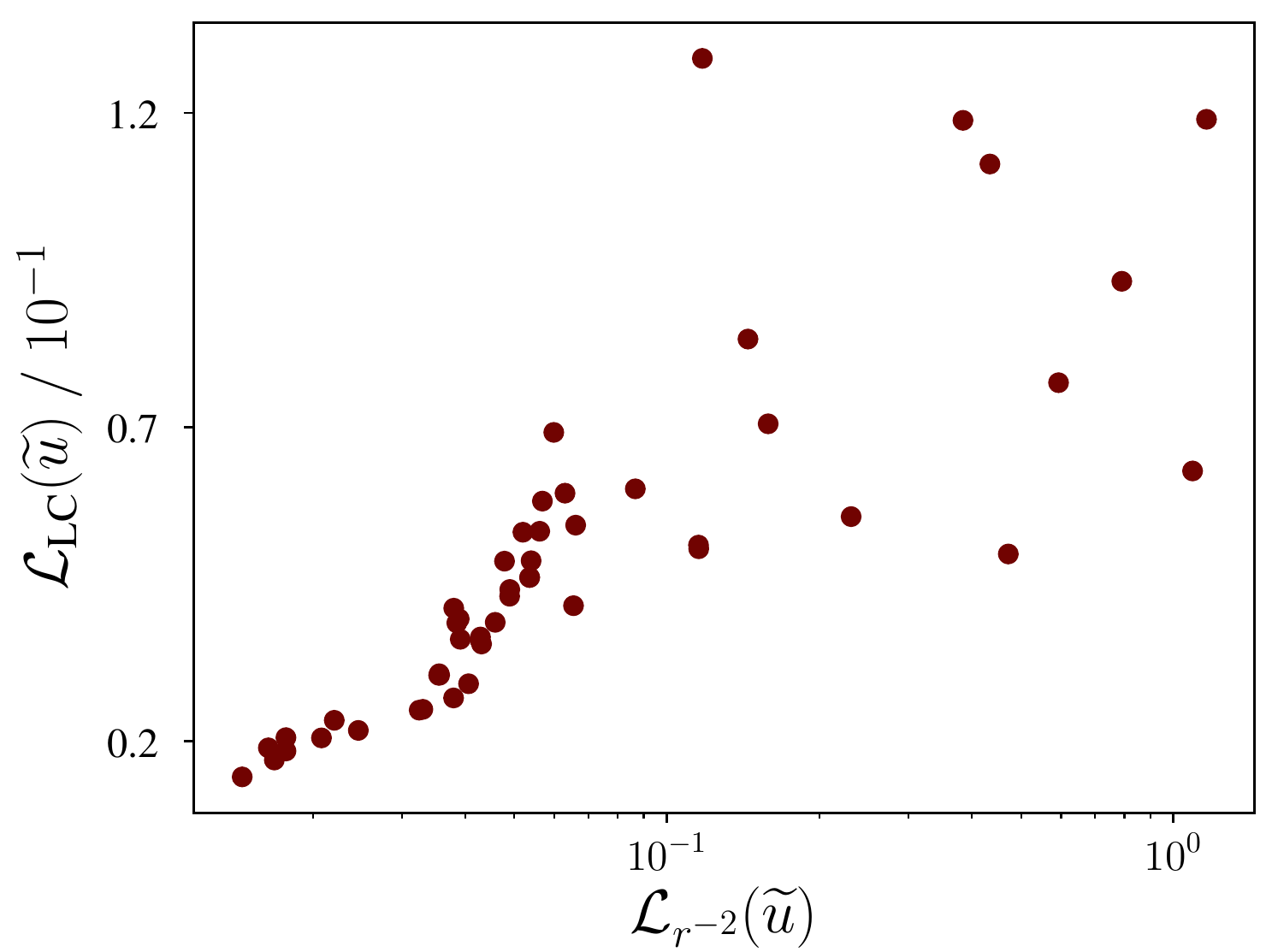}
\caption{Each point represents one of the fifty fittest individuals resulting from the elitist GA. Their coordinates are determined by the corresponding LCLoss and r${}^{-2}$Loss. All LCLosses on the y-axis are given in units of $10^{-1}$. While the plotted distribution is more diffuse at larger r${}^{-2}$Losses, there is a clear correlation between the r${}^{-2}$Loss and the LCLoss at lower losses. For instance, the fittest individual with LCLoss $1.436\times 10^{-2}$ has also the lowest r${}^{-2}$Loss $1.451\times 10^{-2}$.}
\label{fig:LCLossAndR2Loss}
\end{figure}

Based on the broad $95\%$ prediction interval for $\log(r)\gtrsim -3$ in Fig.~\ref{fig:LinearFitAndPredictionIntervals}, we conclude that there are quite a few outliers at larger distances. Therefore, we need to investigate whether these outliers are merely a product of statistics during the GA or if they actually indicate the presence of an alternative LC potential beyond the inverse square potential, which merely could not be reached due to suboptimal crossover and mutation hyperparameters. Let us introduce the inverse-square-loss (r${}^{-2}$Loss) for a transformed potential $\widetilde{\bm{u}}$ as
\begin{equation}
\mathcal{L}_{r^{-2}}(\widetilde{\bm{u}}) = \sqrt{\int_{-12}^0 \! \mathrm{d}\log(r)\left(\widetilde{u}(r)+\frac{\log(r)}{4}\right)^2}.
\end{equation}
Note that $\bm{u}$ is penalized the more its shape deviates from that of an inverse-square potential as in Eq.~\eqref{eq:inverse-square-x1}. This r${}^{-2}$Loss is constructed to vanish when evaluated for $\widetilde{\bm{u}}$ from Eq.~\eqref{eq:transformed-inverse-square-potential}. Fig.~\ref{fig:LCLossAndR2Loss} shows how LCLoss and r${}^{-2}$Loss are related with each other for the fifty fittest individuals. We can observe a strong correlation between the two losses at lower values. Although the point cloud becomes more diffuse at higher r${}^{-2}$Losses, we can still conclude that all outliers have larger LCLosses and are merely of statistical origin. Finally, this also rules out the above hypothesized existence of alternative LC potentials.

\begin{figure}[t]
\includegraphics[width=\textwidth]{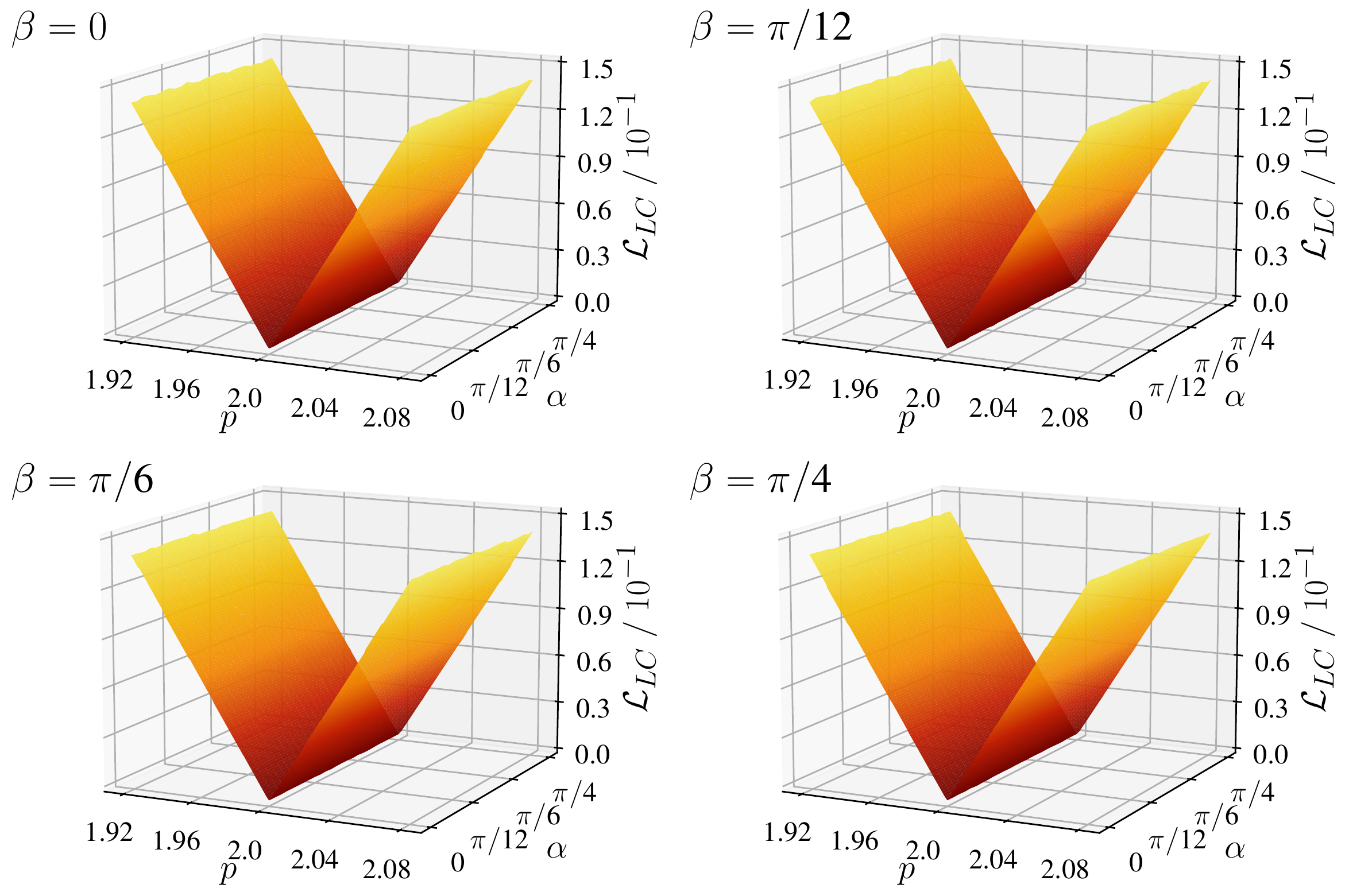}
\caption{Loss surfaces for the LCLoss of $1/r^p$ potentials with exponents $p\in[1.92,2.08]$ and the hyperangle $\alpha\in[0,\pi/4]$ evaluated at four discrete hyperangles $\beta\in\{0, \pi/12, \pi/6, \pi/4\}$. While each individual surface is almost perfectly independent of $\alpha$, the four surfaces also strongly resemble each other, which indicates independence of the hyperangle $\beta$. Since each surface has a distinctive ravine at $p=2$, the inverse-square potential is also the only LC potential at other hyperangles than $\alpha,\beta=0$.}
\label{fig:LCLossSurface}
\end{figure}
So far, calculations of LCLosses have been carried out at specific hyperangles $\alpha=0$ and $\beta=0$. All that remains for us to verify is that the inverse-square potential remains the only LC potential in the four-boson sector for other hyperangle configurations. Similar to Ref.~\cite{arXiv:2111.07820} we observe that at fixed scales $s$ the pole density of $H(R_*,\alpha,\beta)$ is not constant, but a function of the hyperangles. With the definitions of the two hyperangles in Eq.~\eqref{eq:def-hyperangles-x1}, the regime of high-pole densities is located close to the limits $\alpha\to\pi/2$ and $\beta\to\pi/2$. Since we want to avoid misidentifying poles of $H(R_*,\alpha,\beta)$ due to pole densities that are larger than the hyperradial resolution, we cannot inspect the whole hyperangular domain. Instead, we restrict the evaluation of LCLosses for $1/r^p$ potentials to configurations $(p, \alpha)$ with $1.92\leq p \leq 2.08$ and $0 \leq \alpha \leq \pi/4$. The following procedure is the same as in Ref.~\cite{arXiv:2111.07820}. However, the second hyperangle $\beta$, which is introduced by the fourth particle, is an additional degree of freedom, which must be taken into account. For the sake of readability, we restrict it to the four discrete values $\beta\in\{0, \pi/12, \pi/6, \pi/4\}$. Fig.~\ref{fig:LCLossSurface} displays the four corresponding loss surfaces. These are essentially identical, that is independent of $\beta$, independent of $\alpha$ and and all have a distinctive ravine with $\mathcal{L}_\text{LC}\sim\mathcal{O}(10^{-5})$ at the expected exponent $p=2$. Finally, we have verified that the inverse-square potential also minimizes the LCLoss for hyperangle configurations beyond $\alpha,\beta = 0$.

\section{Discussion}
As it has already been shown in Ref.~\cite{arXiv:2111.07820}, the inverse-square potential is the only two-body potential leading to an RG limit cycle in systems of three identical bosons without requiring the two-body $S$-wave scattering length to approach the unitary limit. In this work, we address the question whether that observation can also be continued to the next complicated few-body system, that is the system of four identical bosons interacting exclusively by an attractive, piecewise-constant and finitely-ranged two-body potential. 

The starting point of our search for the desired four-boson RG limit cycles is the corresponding low-energy Faddeev equation in hyperspherical coordinates. By means of a generalized transfer matrix method the individual local solutions can be connected with each other in a continuous and differentiable way, from which the total low-energy four-boson wavefunction follows. Subsequently, we perform a delta-shell regularization, which yields a coupling constant $H(R_*,\alpha,\beta)$ depending on a cutoff hyperradius $R_*$ and the two hyperangles $\alpha,\beta$. For sufficiently deep potentials, numerous poles along the $R_*$-axis emerge for $H(R_*,\alpha,\beta)$. The LCLoss measuring the log-periodicity of poles in $H(R_*,\alpha,\beta)$ is heavily inspired by the LCLoss in Ref.~\cite{arXiv:2111.07820}: By construction, a vanishing LCLoss indicates perfect log-periodicity of $H(R_*,\alpha,\beta)$ and, therefore, discrete scale symmetry or, equivalently, an RG limit cycle.

Using the same boosted ensemble of VAEs as in Ref.~\cite{arXiv:2111.07820}, we initialize populations of singular, attractive, piecewise-constant and finitely-ranged two-body potentials in its low-dimensional latent space. In this context, the LCLoss can also be understood as a fitness parameter for each individual within these populations, which is an important prerequisite for GAs. Applying the same GA as in Ref.~\cite{arXiv:2111.07820} to fifty populations in parallel and, thereby, reducing the search for limit cycles to an optimization problem in latent space, we observe the fittest individuals from each population to accumulate around one specific point in latent space. Again, we identify this point as the inverse-square potential, for which the LCLoss vanishes independent of the considered hyperangle configuration. The existence of alternative LC potentials in the four-boson system is ruled out due to the correlation between the LCLoss and the r${}^{-2}$Loss.

Apart from obvious limitations like the restriction to zero-node solutions and finitely-ranged and piecewise-constant two-body potentials, there are other non-trivial limitations to our results that need to be discussed: The error of the fitted slope in Eq.~\eqref{eq:exponent-x1} is sufficiently small to reliably establish a connection to the inverse-square potential. However, it turns out to exceed the corresponding error on the slope in Ref.~\cite{arXiv:2111.07820} by one order of magnitude. The question how this significantly larger error occurs breaks down to investigating the origin of the responsible outliers. There are only two aspects in which the approach of this work differs from that in Ref.~\cite{arXiv:2111.07820}. Firstly, the scales $s$ on which the ensemble $\mathcal{S}$ is trained are derived by a bisection and not by a grid search as in Ref.~\cite{arXiv:2111.07820}. But since this bisection provides a much finer scale-resolution in comparison to the referred grid search, this cannot serve as a possible explanation. The remaining aspect, which, therefore, somehow needs to explain the larger error, is the presence of a fourth boson itself. Note that the Bessel functions within the zero-node solutions have larger indices than in Ref.~\cite{arXiv:2111.07820}. This entails a tendency to larger node frequencies and, consequently, larger pole frequencies for the coupling constants. It is likely that the LCLoss in this work is less error-tolerant than the LCLoss used for three identical bosons in Ref.~\cite{arXiv:2111.07820}. If this proves to be true, this problem could become more severe for additional particles and poses numerical challenges on analyses of RG limit cycles for $A$-body systems with $A>4$.

During the analysis of loss surfaces for $1/r^p$ potentials, we have only considered hyperangles $\alpha,\beta\in[0,\pi/4]$. Note, that the ensemble $\mathcal{S}$ is only trained to scale potentials to produce pole numbers around the target value $N_\Pi=10^2$ at the specific hyperangles $\alpha=0$ and $\beta=0$. For hyperangles $\alpha,\beta\gtrsim \pi/4$, we observe that this scaling no longer suffices to keep the pole frequency below the hyperradial resolution. Strictly speaking, it would be necessary to train an alternative ensemble $\mathcal{S}^\prime$ to make the regime $\alpha,\beta\in[\pi/4,\pi/2]$ accessible to our loss surfaces. However, the loss surfaces in our restricted hyperangular domain turn out to be independent of both hyperangles. Therefore, we simply assume that this trend continues to larger hyperangles as well.

Although there is much overlap in methodology with Ref.~\cite{arXiv:2111.07820}, the result of this work must be considered separately. Since the results of Ref.~\cite{arXiv:2111.07820} apply to each of the four three-boson subsystems, we arrive at the following, central conclusion: If the three-boson subsystems individually give rise to an RG limit cycle, then an RG limit cycle can be automatically predicted for the entire four-boson system.

First of all, this observation is deeply related to Ref.~\cite{PhysRevA.70.052101}, where it is shown that after a renormalization of the two- and three-body subsystems, tetramer energies already turn out to be independent of the UV-cutoff. If we had found a different LC potential than the inverse-square potential, this would indicate that an additional scale emerges in the four-body system and that a four-body term is required at LO. However, we could rule out the existence of other LC potentials, which we also understand as further validation of Ref.~\cite{PhysRevA.70.052101}.

Secondly, the question arises whether this pattern continues to more complex few-body systems. Considering only systems of identical bosons interacting through one and the same two-body potential beyond the unitary limit, this would imply that RG limit cycles in $A^{\prime\prime}$-body subsystems also cause RG limit cycles to emerge in any higher-order $A^\prime$-body subsystems with $3\leq A^{\prime\prime} < A^\prime \leq A$ and arbitrary $A>4$. 
Within the restricted framework of finitely-ranged and piecewise-constant two-body potentials, the hyperspherical and local zero-node solutions behave like a superposition $\left(C J_x(k_x R)+D Y_x(k_x R)\right)/R^{y}$ for any numbers $A$ of particles. Therefore, a thorough investigation of such generalized wavefunctions with arbitrary $x,y\in\mathbb{R}_+$ and the corresponding coupling constants could shed light on this conjecture.

\section*{Acknowledgements} 
We  acknowledge  partial  financial support by the Deutsche Forschungsgemeinschaft
(DFG, German Research Foundation) and the NSFC through the funds provided  to  the  Sino-German  Collaborative  Research  Center  TRR110  ``Symmetries  and  the  Emergence  of  Structure in  QCD''  (DFG  Project  ID
196253076  -  TRR  110,  NSFCGrant  No.  12070131001).  Support  was also provided
by the Chinese Academy of Sciences (CAS) President's International Fellowship Initiative (PIFI)
(Grant No. 2018DM0034), by  Volkswagen Stiftung  (Grant  No.  93562),  and  by  the  EU Horizon 2020 (Grant No. 824093). Further, this project has received funding from the European Research Council (ERC) under the
European Union's Horizon 2020 research and innovation programme (grant agreement No. 101018170).

\begin{appendix}
\section{Low-energy Faddeev equation in the four-boson sector}\label{app:a}
In a system of four identical bosons with mass $m$, one can identify twelve degrees of freedom, e.g. in the form of the four position vectors $\bm{r}_1,\bm{r}_2,\bm{r}_3,\bm{r}_4$ in coordinate space. Since we are not interested in center-of-mass dynamics, these can be seperated out, which corresponds to a reduction by three degrees of freedom. The Jacobi coordinates
\begin{equation}
\bm{j}_{ba}=\frac{1}{\sqrt{2}}\left(\bm{r}_b-\bm{r}_a\right), \ \bm{j}_{cba}=\sqrt{\frac{2}{3}}\left(\bm{r}_c-\frac{\bm{r}_b+\bm{r}_a}{2}\right), \ \bm{j}_{dcba}=\frac{\sqrt{3}}{2}\left(\bm{r}_d-\frac{\bm{r}_c+\bm{r}_b+\bm{r}_a}{3}\right), \label{eq:jacobi}
\end{equation}
are suitable for expressing the resulting Hamilton operator of relative motion in a mathematically simple manner: With $\Delta_{ba},\Delta_{cba},\Delta_{dcba}$ being the Laplacians with respect to the Jacobi vectors $\bm{j}_{ba},\bm{j}_{cba},\bm{j}_{dcba}$ from Eq.~\eqref{eq:jacobi}, the relative motion kinetic energy operator becomes
\begin{equation}
T = -\frac{1}{2m}(\Delta_{ba}+\Delta_{cba}+\Delta_{dcba}).\label{eq:relative-kinetic-energy}
\end{equation}
Here, $(a,b,c,d)$ is some permutation of $(1,2,3,4)$ if not stated otherwise. The underlying permutation symmetry directly follows from the assumption of all four bosons being identical. Meanwhile, we restrict our analysis to finitely-ranged and isotropic two-body potentials $v(r)$ that only depend on the inter-particle distance $r$. Since there are six different undirected edges between four particles, the resulting potential can be represented as a sum of six different contributions,
\begin{equation}
V(\bm{r}_1,\bm{r}_2,\bm{r}_3,\bm{r}_4) = \sum\limits_{a=1}^3\sum\limits_{b>a}^4 v\left(\sqrt{2}j_{b a}\right).\label{eq:total-potential}
\end{equation}
The resulting Hamiltonian of relative motion is then given as the sum $H=T+V$ of the kinetic energy operator $T$ from Eq.~\eqref{eq:relative-kinetic-energy} and the above potential $V$ from Eq.~\eqref{eq:total-potential}.

For further reducing degrees of freedom, we first need to transform all dependencies on magnitudes of Jacobi coordinates within the Hamiltonian of relative motion to hyperspherical coordinates, see Ref.~\cite{BRAATEN2006259}. These are comprised of the hyperradius
\begin{equation}
R = \sqrt{j_{ba}^2+j_{cba}^2+j_{dcba}^2}
\end{equation}
and, in the case of four particles, two hyperangles
\begin{equation}
\alpha_{cba} = \arcsin\left(\frac{j_{cba}}{\sqrt{j_{ba}^2+j_{cba}^2}}\right), \ \beta_{dcba} = \arcsin\left(\frac{j_{dcba}}{\sqrt{j_{ba}^2+j_{cba}^2+j_{dcba}^2}}\right).
\end{equation}
Therefore, the magnitudes of the Jacobi coordinates can be uniquely determined once all hyperspherical coordinates are known and vice-versa,
\begin{equation}
j_{ba}=R\cos\alpha_{cba}\cos\beta_{dcba},\ j_{cba}=R\sin\alpha_{cba}\cos\beta_{dcba},\ j_{dcba}=R\sin\beta_{dcba}.
\end{equation}
Using the chain-rule, derivatives transform as follows:
\begin{equation}
\frac{\mathrm{d}}{\mathrm{d}j_{ba}} = \cos\alpha_{cba}\cos\beta_{dcba}\frac{\partial}{\partial R}-\frac{1}{R}\frac{\sin\alpha_{cba}}{\cos\beta_{dcba}}\frac{\partial}{\partial \alpha_{cba}}-\frac{1}{R}\cos\alpha_{cba}\sin\beta_{dcba}\frac{\partial}{\partial \beta_{dcba}}\label{eq:chain-1}
\end{equation}
\begin{equation}
\frac{\mathrm{d}}{\mathrm{d}j_{cba}} = \sin\alpha_{cba}\cos\beta_{dcba}\frac{\partial}{\partial R}+\frac{1}{R}\frac{\cos\alpha_{cba}}{\cos\beta_{dcba}}\frac{\partial}{\partial \alpha_{cba}}-\frac{1}{R}\sin\alpha_{cba}\sin\beta_{dcba}\frac{\partial}{\partial \beta_{dcba}}\label{eq:chain-2}
\end{equation}
\begin{equation}
\frac{\mathrm{d}}{\mathrm{d}j_{dcba}} = \sin\beta_{dcba}\frac{\partial}{\partial R}+\frac{1}{R}\cos\beta_{dcba}\frac{\partial}{\partial \beta_{dcba}}\label{eq:chain-3}
\end{equation}
Let $j$, $\Delta$ and $L$ denote the magnitude, Laplacian and, respectively, the angular momentum operator of one of the Jacobi coordinates from Eq.~\eqref{eq:jacobi}. These are related to each other as follows:
\begin{equation}
\Delta=\frac{1}{j^2}\frac{\partial}{\partial j}\left(j^2\frac{\partial}{\partial j}\right)-\frac{L^2}{j^2} \label{eq:laplacian}
\end{equation}
Note that the mentioned coordinate transform requires us to insert either Eq.~\eqref{eq:chain-1}, \eqref{eq:chain-2} or \eqref{eq:chain-3} into Eq.~\eqref{eq:laplacian}, depending on the actual choice of the Jacobi coordinate. Continuing this to each summand in Eq.~\eqref{eq:relative-kinetic-energy}, we obtain the relative kinetic energy operator
\begin{equation}
T=T_R+T_{\alpha}+T_{\beta}+\frac{\Lambda^2}{2mR^2}
\end{equation}
with the hyperradial kinetic energy operator
\begin{equation}
T_R=-\frac{1}{2m}\left(\frac{\partial}{\partial R^2}+\frac{8}{R}\frac{\partial}{\partial R}\right)
\end{equation}
as well as the two hyperangular kinetic energy operators
\begin{equation}
T_{\alpha}=-\frac{1}{2mR^2\cos^2\beta_{dcba}}\left(\frac{\partial}{\partial \alpha_{cba}^2}+4\cot(2\alpha_{cba})\frac{\partial}{\partial \alpha_{cba}}\right) \label{eq:hyperangular-kinetic-alpha}
\end{equation}
and
\begin{equation}
T_{\beta}=-\frac{1}{2mR^2}\left(\frac{\partial}{\partial \beta_{dcba}^2}+(2\cot\beta_{dcba}-5\tan\beta_{dcba})\frac{\partial}{\partial \beta_{dcba}}\right).\label{eq:hyperangular-kinetic-beta}
\end{equation}
Together with the grand angular momentum operator
\begin{equation}
\Lambda^2=\frac{L_{ba}^2}{\cos^2\alpha_{cba}\cos^2\beta_{dcba}}+\frac{L_{cba}^2}{\sin^2\alpha_{cba}\cos^2\beta_{dcba}}+\frac{L_{dcba}^2}{\sin^2\beta_{dcba}},
\end{equation}
these agree with the recursion relation for hyperspherical operators from Ref.~\cite{10.1016/0003-4916(83)90212-9}.
We now decompose the wave function into six Fadeev components, as demonstrated in Ref.~\cite{10.1007/s00601-019-1529-5},
\begin{equation}
\Psi(\bm{r}_1,\bm{r}_2,\bm{r}_3,\bm{r}_4) = \sum\limits_{(a,b,c,d)\in Q} \psi(\bm{j}_{ba},\bm{j}_{cba},\bm{j}_{dcba}). \label{eq:faddeev-decomposition}
\end{equation}
In contrast to the three-particle sector, where only three Faddeev components are required, here the number of six components arises from the fact that there are six undirected edges between four particles. As an index set,
\begin{equation}
Q=\{(1,2,3,4),(1,3,2,4),(1,4,2,3),(2,3,1,4),(2,4,1,3),(3,4,1,2)\}
\end{equation}
labels all of these edges. The key idea of the approach in Eq.~\eqref{eq:faddeev-decomposition} is that the total wavefunction $\Psi$ solves the Schr\"odinger equation,
\begin{equation}
H \Psi(\bm{r}_1,\bm{r}_2,\bm{r}_3,\bm{r}_4) = E \Psi(\bm{r}_1,\bm{r}_2,\bm{r}_3,\bm{r}_4),
\end{equation}
if the Faddeev component $\psi$ satisfies the coupled system of six Faddeev equations
\begin{equation}
\left(T_R+T_{\alpha}+T_{\beta}+\frac{\Lambda^2}{2mR^2}-E\right)\psi(\bm{j}_{ba},\bm{j}_{cba},\bm{j}_{dcba})= -v(\sqrt{2}j_{ba})\Psi(\bm{r}_1,\bm{r}_2,\bm{r}_3,\bm{r}_4). \label{eq:faddeev-equation}
\end{equation}
Note that these six equations differ only by the choice of the indices $(a,b,c,d)\in Q$ in the arguments of the two-body potential $v$ and the Faddeev component $\psi$ itself. In contrast, the indices within the hyperangular kinetic energy operators $T_\alpha,T_\beta$ remain fixed among all six Faddeev equations as shown in Eqs.~\eqref{eq:hyperangular-kinetic-alpha} and \eqref{eq:hyperangular-kinetic-beta}, e.g. $(a,b,c,d)=(1,2,3,4)$.

The harmonic expansion of the Faddeev component $\psi$ is given by
\begin{align}
\psi(\bm{j}_{ba},\bm{j}_{cba},\bm{j}_{dcba})=\sum\limits_{l_{ba},m_{ba}}\sum\limits_{l_{cba},m_{cba}}\sum\limits_{l_{dcba},m_{dcba}}\psi_{l_{ba}m_{ba}; l_{cba}m_{cba}; l_{dcba}m_{dcba}}(R,\alpha_{cba},\beta_{dcba})\notag \\ \times Y_{l_{ba}m_{ba}}\left(\bm{\hat{j}}_{ba}\right)Y_{l_{cba}m_{cba}}\left(\bm{\hat{j}}_{cba}\right)Y_{l_{dcba}m_{dcba}}\left(\bm{\hat{j}}_{dcba}\right).\label{eq:faddeev-component-expansion-1}
\end{align}
Again, we recognize the nine degrees of freedom in terms of the three hyperspherical coordinates $R,\alpha_{cba},\beta_{dcba}$ and three normalized Jacobi coordinates $\bm{\hat{j}}_{ba},\bm{\hat{j}}_{cba},\bm{\hat{j}}_{dcba}$. However, considering the four bosons explicitly in the low-energy regime allows us to drop terms with higher angular momenta, since these are associated with higher energies. Ref.~\cite{NIELSEN2001373} argues that neglecting terms with $l_{ba},l_{cba},l_{dcba}> 0$ is a more accurate approximation in the context of the Faddeev approach than during a direct attempt of solving the Schr\"odinger equation. With the abbreviation $\psi_0=\psi_{00;00;00}$, Eq.~\eqref{eq:faddeev-decomposition} reduces to
\begin{equation}
\Psi(\bm{r}_1,\bm{r}_2,\bm{r}_3,\bm{r}_4) = \sum\limits_{(a,b,c,d)\in Q} \psi_0(R,\alpha_{cba},\beta_{dcba}). \label{eq:faddeev-decomposition-2}
\end{equation}

Let us now specify the indices $(a,b,c,d)=(1,2,3,4)$. We may substitute the Faddeev component in Eq.~\eqref{eq:faddeev-equation} with $\psi_0$, such that the left-hand side now only depends on the hyperradius $R$ and the two hyperangles $\alpha=\alpha_{321}$ and $\beta=\beta_{4321}$. However, there are residual couplings to the other five Faddeev equations due to the remaining contributions to $\Psi$. These manifest in terms of implicit dependencies on other indices $(a^\prime,b^\prime,c^\prime,d^\prime)\in Q\backslash\{(a,b,c,d)\}$ that only appear on the right-hand side. Since the three Jacobi vectors $\bm{\hat{j}}_{ba},\bm{\hat{j}}_{cba},\bm{\hat{j}}_{dcba}$ are a basis of $\mathbb{R}^3$ as long as none of the positions $\bm{r}_1,\bm{r}_2,\bm{r}_3,\bm{r}_4$ coincide with each other, we can understand the remaining hyperangles $\alpha_{c\prime b^\prime a^\prime},\beta_{d^\prime c\prime b^\prime a^\prime}$ as functions on $R,\alpha,\beta$ and the three angles $\theta_1,\theta_2,\theta_3$ between the Jacobi vectors with $\cos\theta_1=\bm{\hat{j}}_{ba}\cdot \bm{\hat{j}}_{cba},\, \cos\theta_2=\bm{\hat{j}}_{ba}\cdot \bm{\hat{j}}_{dcba},\, \cos\theta_3=\bm{\hat{j}}_{cba}\cdot\bm{\hat{j}}_{dcba}$,
\begin{equation}
\alpha_{c^\prime b^\prime a^\prime}=\alpha_{c^\prime b^\prime a^\prime}\left(R,\alpha,\beta;\cos\theta_1,\cos\theta_2,\cos\theta_3\right),
\end{equation}
\begin{equation}
\beta_{d^\prime c^\prime b^\prime a^\prime}=\beta_{d^\prime c^\prime b^\prime a^\prime}\left(R,\alpha,\beta;\cos\theta_1,\cos\theta_2,\cos\theta_3\right).
\end{equation}
Therefore, each $\psi_{0}(R,\alpha_{c^\prime b^\prime a^\prime },\beta_{d^\prime c^\prime b^\prime a^\prime })$ can be expanded in terms of Legendre polynomials, 
\begin{align}
\psi_{0}(R,\alpha_{c^\prime b^\prime a^\prime },\beta_{d^\prime c^\prime b^\prime a^\prime }) = \sum\limits_{x}\sum\limits_{y}\sum\limits_{z} \psi_{xyz}^{(a^\prime b^\prime c^\prime d^\prime )}(R,\alpha_{cba},\beta_{dcba})\notag \\ \times P_x\left(\bm{\hat{j}}_{ba}\cdot \bm{\hat{j}}_{cba}\right)P_y\left(\bm{\hat{j}}_{ba}\cdot \bm{\hat{j}}_{dcba}\right)P_z\left(\bm{\hat{j}}_{cba}\cdot \bm{\hat{j}}_{dcba}\right) \label{eq:faddeev-component-expansion-2}
\end{align}
Due to the orthogonality of Legendre polynomials, the contributions of $\cos\theta_1,\cos\theta_2,\cos\theta_3$ can be easily eliminated by applying the triple integral
\begin{equation}
\frac{1}{2^3}\int_{[-1,+1]^3} \! \mathrm{d}^3\cos\theta = \frac{1}{2^3}\int_{-1}^{+1} \! \mathrm{d}\cos\theta_1\int_{-1}^{+1} \! \mathrm{d}\cos\theta_2\int_{-1}^{+1} \! \mathrm{d}\cos\theta_3
\end{equation}
on both sides of Eq.~\eqref{eq:faddeev-equation}. The remaining Faddeev components on the right-hand side, which have now become independent of $\cos\theta_1,\cos\theta_2,\cos\theta_3$ can be reproduced by a projection operator
\begin{equation}
I = \frac{1}{2^3}\int_{[-1,+1]^3} \! \mathrm{d}^3\cos\theta \int_{\mathbb{R}^2} \! \mathrm{d}^2\omega^\prime \sum\limits_{(a^\prime\!,b^\prime\!,c^\prime,d^\prime)\in Q \backslash \{(1,2,3,4)\}}\delta^{(2)}(\bm{\omega}^\prime-\bm{\omega}_{d^\prime c^\prime b^\prime a^\prime})\label{eq:projection-operator}
\end{equation}
which is applied to $\psi_0(R,\alpha^\prime,\beta^\prime)$, where $\bm{\omega}^\prime=(\alpha^\prime,\beta^\prime)$ and $\bm{\omega}_{d^\prime c^\prime b^\prime a^\prime}=(\alpha_{c^\prime b^\prime a^\prime},\beta_{d^\prime c^\prime b^\prime a^\prime})$. Thereby, the system of six coupled Faddeev equations has collapsed to one single low-energy Faddeev equation for four identical bosons,
\begin{equation}
(T_R+T_\alpha+T_\beta-E)\psi_0(R,\alpha,\beta) = -v(\sqrt{2}R\cos\alpha\cos\beta)\left[\psi_0(R,\alpha,\beta)+(I\psi_0)(R,\alpha,\beta)\right].\label{eq:le-faddeev-equation}
\end{equation}
In this shape, Eq.~\eqref{eq:le-faddeev-equation} strongly resembles the low-energy Faddeev equation for three identical bosons from Ref.~\cite{BRAATEN2006259}.
Due to the restriction to the low-energy sector and the application of the projection operator $I$ from Eq.~\eqref{eq:projection-operator}, the total number of degrees of freedom has finally been reduced to three, which are covered by the hyperradius $R$ as well as the two hyperangles $\alpha$ and $\beta$.

\section{Local solutions}\label{app:b}
Since we intend to make use of the transfer matrix method, see Ref.~\cite{62122}, we are particularly interested in local solutions of the low-energy Faddeev equation, see Eq.~\eqref{eq:le-faddeev-equation}, for four bosons. Therefore, we consider $\psi_0$ at a fixed hyperradius $R$ and hyperangles $\alpha,\beta$ and assume the two-body potential to be constant in vicinity of $r=\sqrt{2}R\cos\alpha\cos\beta$,
\begin{equation}
v(\sqrt{2}R\cos\alpha\cos\beta)=\frac{u}{2m}.
\end{equation}
Although this already seems to be a strong simplification, solving Eq.~\eqref{eq:le-faddeev-equation} still remains extremely challenging. This is especially due to the nature of the projection operator from Eq.~\eqref{eq:projection-operator}. At the cost of restricting ourselves to zero-node solutions in the hyperangular sector, that is assuming a purely hyperradial wavefunction
\begin{equation}
\psi_0(R,\alpha,\beta) = f(R),
\end{equation}
we obtain a vanishing contibution of the hyperangular kinetic energy operators $T_\alpha,T_\beta$ when applied to $f(R)$ and insert
\begin{equation}
(If)(R) = 5 f(R)
\end{equation}
on the right-hand side of Eq.~\eqref{eq:le-faddeev-equation}. Together with
\begin{equation}
k_u^2 = 2mE-6u
\end{equation}
this yields a simple hyperradial equation of motion
\begin{equation}
\left(\frac{\partial}{\partial R^2}+\frac{8}{R}\frac{\partial}{\partial R}+k_u^2\right)f(R)=0,
\end{equation}
which is solved by the following linear combinations of Bessel functions of first and second kind times the monomial $1/R^{7/8}$,
\begin{equation}
f_{k_u}(R) = \frac{1}{R^{7/8}}\left(A^\prime J_{7/2}(k_u R)+B^\prime Y_{7/2}(k_u R)\right).\label{eq:faddeev-radial}
\end{equation}
At this point it must be remembered that $\psi(R,\alpha,\beta) = f_{k_u}(R)$ is merely a Faddeev component and that we still need to infer the total wavefunction $\Psi(R,\alpha,\beta)$ in vicinity of the hyperspherical coordinates $R,\alpha,\beta$ from Eq.~\eqref{eq:faddeev-radial}. For this, we insert $\psi(R,\alpha,\beta) = f_{k_u}(R)$ into each summand of Eq.~\eqref{eq:faddeev-decomposition-2} and obtain the zero-node and low-energy approximation 
\begin{equation}
\Psi_{k_u}(R) = \frac{1}{R^{7/8}}\left(A J_{7/2}(k_u R)+B Y_{7/2}(k_u R)\right)
\end{equation}
with $A=6A^\prime$ and $B=6B^\prime$.

\end{appendix}

\end{document}